# NEARBY Platform: Algorithm for Automated Asteroids Detection in Astronomical Images

Teodor Stefanut, Victor Bacu, Constantin Nandra, Denisa Balasz, Dorian Gorgan
Computer Science Department, Technical University of Cluj-Napoca, Romania

(teodor.stefanut, victor.bacu, constantin.nandra, denisa.copandean, dorian.gorgan}@cs.utcluj.ro

Ovidiu Vaduvescu
Isaac Newton Group of Telescopes (ING), Santa Cruz de la Palma, Canary Islands, Spain
Instituto de Astrofisica de Canarias (IAC), La Laguna, Canary Islands, Spain

ovidiu.vaduvescu@gmail.com

*Abstract*—In the past two decades an increasing interest in discovering Near Earth Objects has been noted in the astronomical community. Dedicated surveys have been operated for data acquisition and processing, resulting in the present discovery of over 18.000 objects that are closer than 30 million miles of Earth. Nevertheless, recent events have shown that there still are many undiscovered asteroids that can be on collision course to Earth. This article presents an original NEO detection algorithm developed in the NEARBY research object, that has been integrated into an automated MOPS processing pipeline aimed at identifying moving space objects based on the blink method. Proposed solution can be considered an approach of Big Data processing and analysis, implementing visual analytics techniques for rapid human data validation.

## I. Introduction

According to recent research aimed at determining the size and orbital distribution of the population of near-Earth asteroids, there are about 1000 NEAs larger than 1 km, up to approximately $7 \times 10^4$ NEAs larger than 100m and near $4 \times 10^8$ smaller objects that have trajectories passing by less than 30 million miles of Earth at certain times [1], [2], [3]. Out of all these, in the present astronomers are tracking about 18.000 objects: more than 8.000 larger than 140m and about 900 larger than 1 km [4], which means that only about 25% of possible hazardous asteroids (larger than 140m) are currently tracked.

Continuous efforts are made both by professional astronomers and volunteers in discovering as much as possible of the remaining objects and on permanently monitoring the already known ones for better trajectory definition and more accurate estimations on potential hazardous activity. Some of the main challenges are related with limited observation time (large areas to scan only in dark nights with clear sky), technological capabilities (data acquisition equipment limitations), available processing time (captured images must be analyzed quickly in order to allow follow-up monitoring in the same night or, the latest, in the following night).

Addressing these problems, a few dedicated surveys for NEA discovery have been organized in the past two decades, which benefit from telescope time, large CCD cameras, specialized IT infrastructure and software for automated data processing. These efforts have significantly increased the number of discovered artefacts and the percentage of followed space objects.

However, observation time and technical limitations prevent a continuous monitoring of the entire sky. Usually, the observation patterns of these surveys are programmed to revisit each section once or twice a month. Complementary observing efforts are made by individual researchers and groups of volunteers, that are using limited telescope time (typically a few hours each night for two to four nights) and smaller telescopes to conduct surveys. One of the biggest challenges these initiatives are facing is data analysis in shortest time possible.

Automated data processing techniques available in the dedicated surveys are usually proprietary implementations and have been specifically developed for the instruments available in the survey itself. Consequently, teams of enthusiasts and external researchers do not have access to these processing pipelines and need to use different tools for identifying NEAs. The most common technique is "blinking", invented by physicist Carl Pulfrich at Carl Zeiss AG in 1904 [5]. This technique relies on multiple astronomical images of the same sky area, captured a few minutes apart. After instrumental errors are removed and optical corrections are applied the images are aligned "on top of each-other" using stars as a reference. By switching rapidly from one image to another, the human observer can notice that stars are remaining stationary (as the images have been aligned based in their position) while moving objects are changing position.

The "Blinking" analysis method performed manually is not efficient when analyzing large captured images (e.g. 32 CCDs / image) or a large number of zones. As an automated approach, the NEARBY platform [6] proposes a MOPS pipeline that applies all corrections to images, extracts all sources and identifies potentially moving objects. As a last step, a human operator validates findings and eliminates any false positives. This article is focused on the moving objects detection algorithm, and discusses main challenges faces in analyzing data and reducing the number of false positives.

This paper is structured as follows: the following section will shortly analyze other initiatives in automated NEAs discovery. In section III we will present an overview of the NEARBY platform while in sections IV, V and VI we will concentrate on the moving objects detection algorithm itself, obtained results and identified limitations. Chapter VI will present conclusions of this paper.

## II. RELATED WORKS

One of the first automated algorithms for asteroids detection has been created in 1992 for the 0.9m Spacewatch telescope from the University of Arizona [7]. Using the automated asteroids detection algorithm, the survey has successfully identified 14 asteroids in the first 10 months.

In article [8] a highly automated moving object detection software package is presented. Proposed approach maintains high detections efficiency while reducing low false-detection rates by using two independent detection algorithms and combining the results. One of the algorithms is based on a wavelet transformation and detections in image space while the other aims to identify moving objects in a similar manner with our approach: analyzing SEXTRACTOR output and recognizing possible trajectories.

One of the most advanced software package that produces automatic asteroid discoveries and identifications from catalogs of transient detections has been developed for the Pan-STARRS survey under the name of Moving Object Processing System (MOPS) [9]. MOPS achieves >99:5% efficiency in producing orbits from a synthetic but realistic population of asteroids whose measurements were simulated for a Pan-STARRS4- class telescope.

In article [10], Copandean et al. propose an automated pipeline prototype for asteroids detection, written in Python under Linux. For images preprocessing and correction, 3rd party astrophysics libraries have been used, while for asteroids detection a custom approach based on SEXTRACTOR output has been implemented. The main steps of the algorithm were focused on: (1) eliminating stars and galaxies based on combined catalogs generated also by SEXTRACTOR; (2) process all remaining objects grouped by their acquisition image and ordered by astronomical seeing conditions; (3) identify trajectories based on the deviation angle between the pivot element (taken from the clearest image) and all potential candidates from different images.

Although used in renowned surveys with great results and proven efficiency, most of the solutions above are proprietary and details about their implementation is not accessible. Our aim is to provide an open-source asteroids detection algorithm that can be easily adapted to different observing instruments (i.e. telescopes) and allows large data sets analysis in almost real-time, in an automated manner and with minimum intervention from human observers.

## III. NEARBY PLATFORM OVERVIEW

As already mentioned, the use of "blinking" method by hand is feasible only for smaller images and shorter surveys (where images are captured only from a few sky areas at a time). For any survey that uses high resolution CCDs and aims to cover extended sky areas over multiple nights, an automated processing pipeline is required. As most of the current automated solutions have been developed specifically for the instruments available in each large survey and are usually proprietary software, NEARBY platform constitutes an accessible solution for individual astronomers and groups of volunteers that would like to analyze large amounts astronomical images in almost real-time.

NEARBY platform's pipeline has been inspired from the Moving Object Processing System (MOPS) used in the Pan-STARRS system. As depicted in Fig. 1, the pipeline accepts raw astronomical images as input and performs all the required corrections and instrumental noise elimination before running the Asteroids detection algorithm. The output of this pipeline is a list of potential asteroids that are presented to a human validator for false positives elimination.

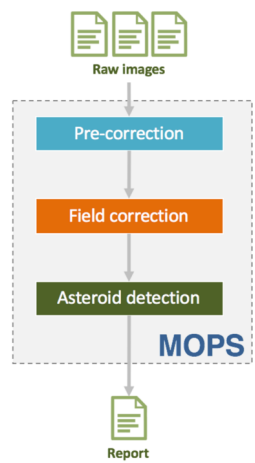

Fig. 1. MOPS modules

NEARBY pipeline has been implemented using Python and the following third-party libraries:

- **IRAF** [11] - a collection of software packages developed by the National Optical Astronomy Observatory (NOAO) aimed to process astronomical images. This library is mainly used for images headers editing, formats conversion (e.g. between FITS and PNG), image cropping.

- **SEXTRACTOR** [12] – process images and extract and build a catalog of astronomical objects

- **SCAMP** [13] – compute shifting function used to correct field distortion

- **SWARP** [14] – resample images based on the shifting functions

Ideally, captured images of the same sky area should be perfectly aligned, the telescope positioning being computed as to compensate for Earth rotation in between two consecutive takes (from a few minutes to few hours). However, in reality this is not the case and he correction and processing pipeline needs to take this into consideration. As a result, when the **Field correction** module finishes execution, all the images inserted into the pipeline have been corrected, resampled and realigned according to the requirements of the "blinking" technique: in all images, stars have the same RA and DEC coordinates, so they appear as stationary for the entire time interval from the first to the last image. In opposition, any asteroid should be represented as an object that changes position in consecutive images.

## IV. ASTEROIDS DETECTION ALGORITHM

After the input images have been corrected and synchronized, SEXTRACTOR module is used to extract all astronomical objects from the images, being stars, noise or asteroids. The output of this phase is a catalog in ASCII format for each input image, that includes specific information about the identified sources: position in RA and DEC coordinates, apparent magnitude, elliptical form etc. The only link between the identified objects is represented by the image they have been found in. No other contextual information (e.g. relative positioning in catalog, type of source – star, galaxy, asteroid …) is provided.

Obtained SEXTRACTOR catalogs are then processed by the NEARBY Asteroids Detection Algorithm which has as main purpose to identify asteroids trajectories from the elements detected by SEXTRACTOR in the processed images. As input the algorithm receives four types of information:

1. list of all sources identified by SEXTRACTOR (in the form of the aforementioned catalogs);
2. the exact observation time of each image (in Julian Date format);
3. the exposure time for each image (typically 60 seconds);
4. parameters that describe specific attributes for the survey and the asteroids of interest:
   - **pixel scale** – the number of arc minutes covered by each pixel in the image;
   - **minimum and maximum speed ($\mu_{min}$, $\mu_{max}$)** – the speed limits measured in arc seconds / minute for the moving objects of interest; This value should be coordinated with the survey direction and with the time interval between two consecutive images;
   - **maximum allowed speed variation ($\Delta\mu_{max}$)** – the maximum variation that is accepted between different parts of the same trajectory (influenced by velocity and positioning error);
   - **maximum allowed positioning error ($\varepsilon_{max}$)** – the minimum distance between two objects in order to be considered separate entities.

As output, the algorithm generates identified trajectories in MPC [15] format and required image-space coordinates for creating thumbnails of the NEAs candidates.

Our approach assumes that an asteroid, during the entire observation window (from the first to the last image), moves with constant speed and on a trajectory that is perceived as being linear. Searching through all the sources identified by SEXTRATOR, the detection algorithm organizes them into potential trajectories. The main challenge of this approach is generated by the fact that sources positioning has an error of 0.7 arc seconds, which is often comparable with the distance traveled by the asteroid between two consecutive images. For some cases, this enforces relaxations of the speed and trajectory angle computations which, in turn, favors the false positives.

The main algorithm steps are as follows:

### A. Remove fixed objects

In order to reduce the number of elements that can generate false positives, the first step of the algorithm concentrates on eliminating all sources that are fixed in time according to the RA and DEC coordinates. Most likely these represent stars or galaxies and should not be considered when looking for asteroids trajectories. Are considered to be fixed all the sources that are positioned at a distance shorter than the **maximum allowed positioning error** value in two or more images, no matter if they are consecutive or not.

For specific cases, where asteroids are moving very slowly and the time interval between images is short in comparison, the condition should be modified to consider as static sources only the detections that are positioned at a distance shorter than the **maximum allowed positioning error** value in all images.

### B. Establish speed intervals

NEAs velocity can vary significantly, typically from values of 0.05 to 10 arc seconds / minute. In many of the tested astronomical images, this large variation favors the detection of multiple trajectories for the same source, thus creating many potential false positives trajectories, especially in the initialization phase. To overcome this problem, we have adopted a speed interval partitioning approach that automatically splits the interval defined by the minimum and maximum speed values specified as parameters in 0.5 arc seconds / minute intervals.

For each of these intervals, we are running the discovery algorithms and try to identify all moving objects that meet current velocity conditions. All the sources that have been included into trajectories will not be considered for the next computation phases, thus reducing the number of false positives and the possibility to consider the same object for multiple trajectories.

### C. Initialize possible trajectories

In the third step of the algorithm, all sources are ordered according to the acquisition time and the clearest position is chosen as a pivot (i.e. starting point). The pivot is selected based on the astronomical seeing conditions (FWHM) [16] which is one of the parameters retrieved from SEXTRACTOR catalogs. This ensures that the number of possible detections is maximized, as the chances for the asteroids being visible in this image are the highest.

Each of the sources from the pivot image are then paired with all the sources from the previous (or the following) chronological image, keeping for later processing only the pairs that:

- do not have any common positions with other pairs: this condition allows us to eliminate many false positives from the start, as two or more real and independent trajectories, with similar velocities, are extremely rare close enough to share one or more sources;
- by the computed traveled distance indicate a speed that is within the current speed interval.

In this step we have very few information that can be used to filter out false positive pairs. Any two objects from the pivot and another image that meet the velocity conditions are a very good candidate for an initial trajectory.

### D. Trajectories development

After all the initial trajectories have been created, each of the remaining images are processed in chronological order. From each image a maximum of one new source will be added to each trajectory, selected as being the one that has the minimum direction angle deviation and minimum speed variation compared to the already existing elements in trajectory. Because of the positioning errors mentioned above (0.7 arc seconds), these conditions cannot be very restrictive, facilitating high numbers of false positives, especially for trajectories composed of only three sources.

In order to select the best match between each trajectory and each of the available object in the current processed image, a cost function is computed:

$$f = 0.5 * \left(1 - \frac{\|\mu_t - \mu_c\|}{\Delta\mu_{max}}\right) + 0.5 * \left(1 - \frac{\|\alpha_t - \alpha_c\|}{\Delta\alpha_{max}}\right)$$

where:

$\mu_t$ – the average speed on the existing trajectory segments;

$\mu_c$ – the speed computed on the candidate segment for this trajectory, if the current object would be selected for inclusion in the trajectory;

$\Delta\mu_{max}$ – the value of the parameter **maximum allowed speed variation**;

$\alpha_t$ – the average angle between the existing trajectory segments and the chosen reference;

$\alpha_c$ – the angle of the candidate segment for this trajectory, if the current object would be selected for inclusion in the trajectory

$\Delta\alpha_{max}$ – the maximum deviation angle accepted when searching for candidates, computed dynamically as:

$$\Delta\alpha_{max} = \sin^{-1}\frac{2 * \varepsilon_{max}}{L_t}$$

where:

$\varepsilon_{max}$ – the value of the parameter **maximum allowed positioning error**;

$L_t$ – the length of the already defined trajectory;

As can be seen from the definition of $\Delta\alpha_{max}$, it is assumed that the distance traveled by the asteroid between two consecutive image captures is larger than the double of the **maximum allowed positioning error** value.

## V. TESTING RESULTS

As preliminary testing, the algorithm has been run on 16 sets of images captured with the 2.54m INT telescope from La Palma, Spain, which has 4 thinned EEV 2kx4k CCDs and covers a sky area of 34.2 arc minutes. The automated results have been compared to the results obtained by human observers, with the following outcome:

TABLE I. PRELIMINARY TESTING RESULTS

| Field | NEAs found by observer | Validated detections | Success rate |
| --- | --- | --- | --- |
| Field 1 | 39 | 32 | 82% |
| Field 2 | 33 | 24 | 73% |
| Field 3 | 37 | 26 | 70% |
| Field 4 | 33 | 26 | 79% |
| Field 5 | 28 | 25 | 89% |
| Field 6 | 24 | 21 | 88% |
| Field 7 | 25 | 23 | 92% |
| Field 8 | 33 | 28 | 85% |
| Field 9 | 36 | 25 | 69% |
| Field 10 | 24 | 18 | 75% |
| Field 11 | 28 | 24 | 86% |
| Field 12 | 28 | 22 | 79% |
| Field 13 | 25 | 19 | 76% |
| Field 14 | 32 | 27 | 84% |
| Field 15 | 21 | 18 | 86% |
| Field 16 | 33 | 28 | 85% |

As it can be seen in Table 1, the success rate of the proposed algorithm is promising. It is important to underline the fact that these results have been obtained only based on the SEXTRACTOR output and making use of position coordinates only (RA and DEC). No metadata information like position of stars and galaxies or list of known asteroids that should cross through the analyzed field at the observation time are currently used. All the detections rely solely on the captured images.

## VI. CURRENT LIMITATIONS OF THE ALGORITHM

As already mentioned, the asteroids detection algorithm relies on the objects extracted from images by the SEXTRACTOR library. As the performed analysis is based solely on the images, some of the reported objects are erroneous and favor the detection of false positives.

For example, in Figure 2 we can see an object, which is actually a star, crossing a bad-pixel area of the telescope's CCD. The three images are presented chronologically from left to right and give a false impression of a moving object (taking the bad-pixel lines as a reference). At the same time, as the size of the object is changing with time, SEXTRACTOR analysis of the image will report two objects for a), one object at a specific position for b) and one object, with a slightly different position than the one from b) for c).

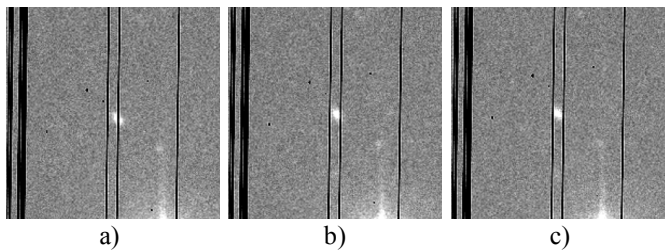

a) b) c)

Fig. 2. MOPS modules

Due to these positioning errors, reported objects have been considered as a valid trajectory by the asteroids detection algorithm. This type of errors can be relatively easily avoided if more information is added into the algorithm, like stars positioning.

## VII. CONCLUSIONS

The automated analysis of astronomical images reduces significantly the time and human resources necessary for NEAs discovery and continuous monitoring. Large datasets can be processed in almost real-time and actions for follow-up on new discoveries or on confirmation of already known asteroids can be taken promptly. However, most of the current automated astronomical images processing solutions are proprietary or have been developed for specific instruments, so they are no suitable to be used by individual astronomers from other observatories or by volunteers and astronomy enthusiasts.

The work presented in this article aims to provide an open-source asteroids detection algorithm that can be easily adapted to different observing instruments (i.e. telescopes) and allows large data sets analysis in almost real-time, in an automated manner and with minimum intervention from human observers. Results obtained so far are promising and current implementation has been successfully integrated in the processing pipeline of the NEARBY Platform.

## ACKNOWLEDGMENT

This research is supported by ROSA (Romanian Space Agency) through the Contract CDI-STAR 192/2017, NEARBY - Visual Analysis of Multidimensional Astrophysics Data for Moving Objects Detection.

## REFERENCES

[1] P. Tricarico, "The near-Earth asteroid population from two decades of observations", Icarus 284, 2017, pp. 416–423

[2] M. Granvik, A. Morbidelli, R. Jedicke, B. Bolin, W.F. Bottke, E. Beshore, D. Vokrouhlický, M. Delbò, P. Michel, "Super-catastrophic disruption of asteroids at small perihelion distances" in Nature 530, 2016, pp. 303–306.

[3] A.W. Harris, G. D'Abramo, "The population of near-earth asteroids", Icarus 257, 2015, pp. 302–312.

[4] NASA Jet Propulsion Laboratory, "Discovery Statistics", [Online] https://cneos.jpl.nasa.gov/stats/totals.html

[5] Zeiss inventions, [Online] http://www.zeiss.com/corporate/en_de/events/international-year-of-light/optical-technologies.html

[6] V. Bacu, A. Sabou, T. Stefanut, D. Gorgan, and O. Vaduvescu, "NEARBY Platform for Detecting Asteroids in Astronomical Images Using Cloud-based Containerized Applications", not published

[7] J. V. Scotti, T. Gehrels, and D. L. Rabinowitz, "Automated detection of asteroids in real-time with the Spacewatch telescope," in Asteroids, Comets, Meteors 1991, A. W. Harris and E. Bowell, Eds., Dec. 1992.

[8] J.-M. Petit, M. Holman, H. Scholl, J. Kavelaars, and B. Gladman, "A highly automated moving object detection package," Monthly Notices of the Royal Astronomical Society, vol. 347, no. 2, pp. 471–480, 2004. [Online]. Available: http://dx.doi.org/10.1111/j.1365-2966.2004.07217.x

[9] L. Denneau, R. Jedicke, T. Grav, M. Granvik, J. Kubica, A. Milani, P. Vere, R. Wainscoat, D. Chang, F. Pierfederici, N. Kaiser, K. C. Chambers, J. N. Heasley, E. A. Magnier, P. A. Price, J. Myers, J. Kleyna, H. Hsieh, D. Farnocchia, C. Waters, W. H. Sweeney, D. Green, B. Bolin, W. S. Burgett, J. S. Morgan, J. L. Tonry, K. W. Hodapp, S. Chastel, S. Chesley, A. Fitzsimmons, M. Holman, T. Spahr, D. Tholen, G. V. Williams, S. Abe, J. D. Armstrong, T. H. Bressi, R. Holmes, T. Lister, R. S. McMillan, M. Micheli, E. V. Ryan, W. H. Ryan, and J. V. Scotti, "The pan-starrs moving object processing system," Publications of the Astronomical Society of the Pacific, vol. 125, no. 926, pp. 357–395, 2013. [Online]. Available: http://www.jstor.org/stable/10.1086/670337

[10] D. Copandean, O. Vaduvescu, and D. Gorgan, "Automated prototype for asteroids detection," in 2017 13th IEEE International Conference on Intelligent Computer Communication and Processing (ICCP), Sept 2017, pp. 377–382.

[11] D. Tody, "The IRAF Data Reduction and Analysis System," in Instrumentation in astronomy VI, vol. 627, Jan. 1986, p. 733.

[12] E. Bertin and S. Arnouts, "SExtractor: Software for source extraction." vol. 117, pp. 393–404, Jun. 1996.

[13] E. Bertin, "Automatic Astrometric and Photometric Calibration with SCAMP," in Astronomical Data Analysis Software and Systems XV, ser. Astronomical Society of the Pacific Conference Series, C. Gabriel, C. Arviset, D. Ponz, and S. Enrique, Eds., vol. 351, Jul. 2006, p. 112.

[14] ——, "SWarp: Resampling and Co-adding FITS Images Together," Astrophysics Source Code Library, Oct. 2010.

[15] Minor Planet Center, Format For Optical Astrometric Observations Of Comets, Minor Planets and Natural Satellites, [Online], https://www.minorplanetcenter.net/iau/info/OpticalObs.html

[16] FWHM explained, [Online], https://www.noao.edu/image_gallery/text/fwhm.html